\documentclass{llncs}

\usepackage{todonotes}
\usepackage{multirow}
\usepackage{listings}
\usepackage{color}
\usepackage{wrapfig}
\usepackage{hyperref}
\usepackage{breakurl}

\usepackage[font=scriptsize,skip=0pt]{caption}

\begin{document}

\title{Model Checking Clinical Decision Support Systems Using SMT}
\titlerunning{Model Checking Clinical Decision Support Systems Using SMT}  
%
\author{Mohammad Hekmatnejad
\inst{1} 
\and Andrew M. Simms
\inst{2,}
\inst{3}
\and Georgios Fainekos
\inst{1} 
}
\authorrunning{Mohammad Hekmatnejad et al.} 
\tocauthor{Mohammad Hekmatnejad, Georgios Fainekos
}
\institute{${^1}$Arizona State University, Tempe AZ 85281, USA,\\
\email{\{mhekmatn,fainekos\}@asu.edu}\\
${^2}$Cognitive Medical Systems, San Diego CA 92121, USA,\\
${^3}$University of Washington, Seattle WA 98195, USA,\\
\email{\{asimms\}@cognitivemedicine.com}
}

\maketitle              

\begin{abstract}
Individual clinical Knowledge Artifacts (KA) are designed to be used in Clinical Decision Support (CDS) systems at the point of care for delivery of safe, evidence-based care in modern healthcare systems. For formal authoring of a KA, syntax verification and validation is guaranteed by the grammar. However, there are no methods for semantic verification. Any semantic fallacy may lead to rejection of the outcomes by care providers. As a first step toward solving this problem, we present a framework for translating the logical segments of KAs into Satisfiability Modulo Theory (SMT) models. We present the effectiveness and efficiency of our work by automatically translating the logic fragment of publicly available KAs and verifying them using Z3 SMT solver.
\end{abstract}

\section{Introduction}
There are substantial ongoing efforts within the healthcare domain looking to improve healthcare quality by helping patients and providers make better decisions whilst avoiding poor ones. Clinical Decision Support (CDS) systems that help patients track, monitor and optimize the care provided to them, or tools that assist providers in analyzing diagnostic data and developing therapeutic plans are among the goals of current research \cite{greenes2011clinical}. 
Evaluating the interaction between such systems, their users, and actual/intended outcomes is an area of particular interest not only to system developers, but also to regulatory organizations such as the FDA, which is responsible for ensuring patient safety. Whether embedded as control logic within a medical device, or deployed as a service component in server, individual CDS knowledge artifacts (models) are often developed in order to optimize health outcomes and efficiency within the context of a specific care process. 

Model checking methods, which were developed for software and hardware verification, have the potential to address the complexity of the aforementioned problem. Model checking is a method by which a modeled system can be validated for compliance against a set of pre-defined specifications. 
Domain knowledge, such as knowledge artifacts incorporated into a software system, is also an artifact that can be model checked. Indeed, any software architected around a corpus of rapidly evolving domain knowledge creates numerous possibilities for model inconsistencies to arise. 
When such artifacts are bundled into more complex systems, or reused in settings different from those assumed by the original scenario, the resulting artifact must be carefully evaluated to ensure that it is still logically and clinically suitable for the new context of care. We propose to utilize Satisfiability Modulo Theory (SMT) based model checking as an effective approach for evaluating KAs. 

{\bf Related work:}
Among all different research lines for the use of Artificial Intelligence (AI) in medicine, there are notable works dedicated to verification of medical device systems \cite{murugesan2013compositional,lee2006high}, model checking Clinical Guidelines (CG) and Clinical Pathways \cite{sciavicco2009quality}.
Formal verification of CGs is not a new research area, and some works such as in \cite{perez2010authoring} modeled the content of CG and verified their properties using different techniques. 
While most of the previous works had to model human readable CGs, we work with some machine readable, standardized and formally defined evidenced base knowledge artifacts. 
Here, we are interested in verifying the satisfiability of formalized logical statements in KAs.     

There exist successful prior arts on the verification of other classes of rule-based systems using SMT verification \cite{vannucchi2017symbolic}, and there already exist methods for computing the minimal set of inconsistent SMT formulas \cite{cimatti2011computing}, which justify our approach.
SMT based model checking has been used for a long history of success in both academia and industry \cite{de2011satisfiability}, and we use it in validating KAs.

\section{Problem}
During the knowledge authoring phase, knowledge engineers and designers translate and bundle clinical knowledge components into KAs. The resulting artifact must be carefully evaluated to ensure that it is logically consistent and clinically suitable for the context of care. From a knowledge management and governance perspective this task becomes exponentially onerous as the number of combinatorial possibilities increases.
Knowledge artifacts can be checked for inconsistencies and conflicts during the authoring phase by using established model verification techniques and tools. For example, a KA about heart failure, may have conditions attached to specific actions so that if they become satisfiable, then their corresponding actions, such as prescribing specific medication orders or procedures can be executed.  These conditions usually are complicated logical expressions that are translated from English narrative to a formal language. There is always a chance that the actual narrative logic or its translated formalization (ELM expression logics) is not sound.

We propose a framework that exploits SMT solvers \cite{de2008z3} to analyze satisfiability of some specifications $\varphi$ (given by domain experts) and expression logics $\mathcal{L}$ of CDS KAs by translating them into SMT formulas $\mathcal{M}$.
If no instance model was attainable, then it means the expression logic is not satisfiable, and therefore unsatisfiable expressions are detectable.
Our proposed solution effectively and efficiently analyzes the embedded logics of CDS KAs.

\noindent

\textbf{Problem Formal Definition} Given a  knowledge artifact $\mathcal{K}$, we are interested in translating expression logics $\mathcal{L}$ embedded in $\mathcal{K}$ into an equivalent formula $\mathcal{M}$ to check for existence of an instance model $\mathcal{S}$, such that $\mathcal{S}$ satisfies $\mathcal{M}$ with respect to a given specification formula $\varphi$.

\section{Preliminaries}


\textbf{Satisfiability Modulo Theories (SMT):} The SMT problem is checking if a given closed logical formula $\varphi$ is \textit{satisfiable} with respect to some background theory $\mathcal{T}$ which restricts the range of used symbols in $\varphi$. In other words, the SMT problem for $\varphi$ and $\mathcal{T}$ is about existence of models of $\mathcal{T}$ that satisfy the formula $\varphi$ \cite{barrett2010smt}.
An \textit{SMT solver} is a software that implements a procedure for satisfiability modulo some given theory. SMT solvers come with different underlying logics, background theories, input formulas and interfaces. 
In this paper, we use the high-performance SMT Solver Z3 that supports all the theories that we need for modeling such as empty theory, linear arithmetic, nonlinear arithmetic, bit-vectors, arrays, data-types, quantifiers and strings \cite{de2008z3}.

 
\noindent
\textbf{CDS Knowledge Artifacts:} 
We will consider CDS KAs based on the HL7 Knowledge Artifact Specification, which is a XML based container for representing clinical knowledge \cite{knart2015}.
The Clinical Quality Language (CQL) \cite{cql2017} is to represent procedural logic and functions within a KA. CQL is written in an XML format called Expression Logical Model (ELM) when written inside a KA. 
KAs are categorized based on their application into three types:
 Event Condition Action (ECA) Rules, Documentation Templates, and Order Sets. Beside their different types, all KAs use the same components as their building blocks. 
In a KA, the only executable components are actions, and in order to see if they can be executed during run-time, we should check if their control conditions evaluate to $true$. In the next section, we assume that the goal is to check if all the underlying Expression Logics represented in ELM are satisfiable despite to coming data in the executing time.

\section{Example of Translating and Verifying A Sample KA}
In this section, we only show some ELM operators and their equivalent SMT in
examples. We are going to use an OS for ``heart failure admission to medical/surgical unit'' as a running example mainly because it has a simple expression logic with the least number of medical terms. This OS has only one condition attached to an action group that has one simple action. Here, we are not focusing on finding contradictions in actions or the contradictions that may occur as a result of executing them as those are out of scope of this paper.
In List \ref{lst:elm:logic}, the Expression Logic of the OS is depicted. It can be noticed that there is a logical $AND$ between two expressions, one is an equality expression and the other is a sequence of logical $NOT$ and $Exist$ operators. This condition in English means \textit{``If the age of patient (evaluated in years) is greater than or equal to 18, and the patient has no history of adverse reaction to ACE inhibitors, then$\ldots$''}. 
We state the equivalent SMT code of the logic in the List \ref{lst:smt:code}. There is an assertion in line 6 that is semantically equivalent to the aforementioned logic in List \ref{lst:elm:logic} as all the ELM operators have equivalent operators in SMT language. We defined the $Exist$ operator as a function in line 4, and for the variables, $AdverseReactionToACEInhibitors$ and $PatientAgeInYears$ are constants of types $List$ $of$ $AdverseEvent$ and $Integer$, respectively. Also, $AdverseEvent$ is not a primitive data type nor a complex one; therefore, it is declared as a sort in SMT. We do not go into details of how these types are decided in the translator, we just mention that the translator extracted them using other sections of KA such as external data and expressions. 
\lstset{
language=xml,
tabsize=3,
caption=Sample ELM expression logic,
label=code:sample,
frame=shadowbox,
rulesepcolor=\color{gray},
xleftmargin=20pt,
framexleftmargin=15pt,
keywordstyle=\color{blue}\bf,
commentstyle=\color{OliveGreen},
stringstyle=\color{red},
numbers=left,
numberstyle=\scriptsize,
numbersep=5pt,
breaklines=true,
showstringspaces=false,
basicstyle=\scriptsize,
emph={conditions,condition,logic},emphstyle={\color{magenta}}}
\begin{lstlisting}[label={lst:elm:logic}]
<logic xsi:type="elm:And">
    <elm:operand xsi:type="elm:GreaterOrEqual">
        <elm:operand xsi:type="elm:ExpressionRef" name="PatientAgeInYears" />
        <elm:operand xsi:type="elm:Literal" valueType="t:Integer" value="18" />
    </elm:operand>
    <elm:operand xsi:type="elm:Not">
        <elm:operand xsi:type="elm:Exists">
            <elm:operand xsi:type="elm:ExpressionRef" name="AdverseReactionToACEInhibitors" />
        </elm:operand> </elm:operand> </logic>
\end{lstlisting}
\vspace{-10pt}
\begin{lstlisting}[basicstyle=\scriptsize,caption=SMT Code equivalent for the example's ELM expression in List \ref{lst:elm:logic},label={lst:smt:code}]
(declare-sort AdverseEvent)
(declare-const AdverseReactionToACEInhibitors (List AdverseEvent))
(declare-const PatientAgeInYears Int)
(define-fun elm_exists ((lst (List AdverseEvent))) Bool
	(ite (exists ((x AdverseEvent)) (= x (head lst))) true false))
(assert (= true (and (>= PatientAgeInYears 18) (not (elm_exists AdverseReactionToACEInhibitors)))))
\end{lstlisting}
For the sake of simplicity, one can check that the logic in List \ref{lst:elm:logic}, can be represented as \textit{(And ($>=$ PatientAgeInYears 18) (Not (Exists AdverseReactionToACEInhibitors)))}. This is almost the same code as is stated in line 6 of List \ref{lst:smt:code}, just with an extra SMT assertion with template ``(assert (= true \textit{(ELM expression logic})))'' to check if the logic statement is satisfiable. 

Note, here we chose a simple use-case to describe the problem and solution, but KAs can have large and complicated embedded logics each with tenths of operators and symbols, such as logics that describe identification of Sepsis and Systemic Inflammatory Response Syndrome (SIRS) ICU \cite{kaukonen2015systemic,pittet1995systemic}. 

We discovered one KA with no satisfiable model from HL7 CDS KA release 1.3\footnote{\scriptsize\url{https://github.com/cqframework/knartwork/tree/master/examples/hl7-cds-ka-r1.3}\normalsize}. This is an ECA Rule 
\footnote{\scriptsize Lines 404-418 at \url{https://cpslab.assembla.com/spaces/cqlverifier/git/source/master/src/main/resources/xml/KNART/ECA-03.xml}\normalsize} 
with a condition in which there is an inequality statement around patient's age informally stated as $(>=$ $18$ $PatientAge)$ $And$ $(<=$ $50$ $PatientAge)$, which is logically a wrong statement. In the SMT translation of the code, SMT solver reported the rule (tagged by name ``assertion-1'') as part of the unsatisfiable core of the code. We added $set-option :produce-unsat-cores$ $true)$ as a Z3 configuration command to force the solver to detect the unsatisfiable assertions, and used $(get-unsat-core)$ for reporting them. 

\section{Experimental Results}

The concept of CDS KAs is new even to the CDS community. Therefore, there are not many KAs available for testing.
In the Table \ref{tbl:performance}, the execution time for seven of the publicly available KAs is presented. 

\begin{wraptable}{r}{4.8cm}
	\vspace{-40pt}
\begin{center}
\resizebox{.4\textwidth}{!}{
\begin{tabular}{ |c|c|c|c|c|c| } 
\hline
\textbf{KA Name} & \textbf{Expr} & \textbf{Oper} & \textbf{Prep} & \textbf{Tran} & \textbf{Solv} \\
\hline
OS-01 & 6 & 31 & 3406 & 26 & 125 \\ 
ECA-01 & 15 & 83 & 3885 & 79 & 148 \\ 
ECA-02 & 19 & 169 & 4128 & 92 & 122 \\
\textbf{ECA-03} & 9 & 76 & 3946 & 64 & 149 \\
ECA-04 & 6 & 36 & 3668 & 47 & 117 \\ 
DT-01 & 1 & 4 & 3228 & 28 & 180 \\
DT-02 & 3 & 13 & 4135 & 30 & 105 \\ 
\hline
\end{tabular}
}
\caption{Running time in milliseconds. All the cases except the \textbf{ECA-03} were satisfiable.}\label{tbl:performance}
\end{center}
	\vspace{-35pt}
\end{wraptable}

Our framework was tested on Mac with the following specifications: 2.6 GHz Intel Core i5 CPU, 16 GB RAM, Z3 Java SMT Solver 4.5, and JDK-8. As it is shown in the Table \ref{tbl:performance}, the verification times (solving column) are efficient for our use-cases because of small size of the models (only conditions). Note that if we translate and incorporate the other elements of a KA, such as actions and behaviors, then based on our experiments, the execution time is potentially going to rise, but we expect it to remain feasible for quantitative model solvers.

\begin{wraptable}{r}{4.8cm}
	\vspace{-40pt}
\begin{center}
\resizebox{.4\textwidth}{!}{
\begin{tabular}{ |l|c|c| } 
\hline
\textbf{Operators} & \textbf{Example} & \textbf{Support} \\
\hline
Logical & And & completely  \\ 
Mathematical & Div & completely  \\
Equality & $>=$ & completely  \\ 
String & startsWith & completely  \\
List & exists & partially  \\ 
Interval & in & partially  \\ 
Time & diffBetween & scarcely  \\ 
Miscellaneous & isTrue & scarcely \\
Aggregation & count & none \\
\hline
\end{tabular}
}
	\caption{Supported operators by their categories.}\label{tbl:operators}
\end{center}
	\vspace{-36pt}
\end{wraptable}

In Table \ref{tbl:operators}, some of the ELM operators that our framework supports for one-to-one translation to SMT are listed by their categories and support coverage.
Note that in Table \ref{tbl:performance}, the higher execution times in the ``Preparing'' column are because of loading all the schema files, verifying artifacts against them, and then unmarshalling them into Java instance objects. 
Our translation tool is currently in a beta version and publicly available at:
\\
\centerline{\url{https://cpslab.assembla.com/spaces/cqlverifier}} \\
under GNU GPLv3 license.

\section{Results and Future Work}
The preliminary results reported here revealed that even KAs with simple logics may have fallacies in them, which need to be fixed in the knowledge authoring phase. For example, ``ECA-03'' in Table \ref{tbl:performance} has two inequality statements about patient's age which are contradictory. Second, we found out that the current HL7 KNART specification does not support range constraints on variables explicitly. For example, while the age of a patient is considered as an $Integer$ variable, there is no assumption regarding the valid age of a live person. 

We intend to apply our tools to a larger and more complex set of knowledge artifacts currently under development that cover multiple clinical specialties, and a broad range of complexity.  Further, we plan to extend the translation capability to the complete definition of ELM expressions described in the standard.

\medskip\medskip
\noindent
\textbf{Acknowledgements} The authors thank Cognitive Medical Systems, inc. for their support and access to their database of KNART artifacts. This work was supported in part by the NSF I/UCRC Center for Embedded Systems and from NSF grant $\#1361926$.

\bibliographystyle{splncs03}

\end{document}